\documentstyle[11pt,aaspp4]{article}

\lefthead{}
\righthead{Stellar Parameters in StH$\alpha$190}

\begin{document}

\title{The Stellar Parameters and Evolutionary State of the Primary in the
d'--Symbiotic System StH$\alpha$190$^{1,2}$}

\author{Verne V. Smith} 
\affil{Department of Physics, University of Texas at El Paso, El Paso, TX 
79968 USA}

\author{and}

\author{Claudio B. Pereira}
\affil{Observatorio Nacional, Rua General Jos\'e Cristino 77, CEP 20921-400,
Rio de Janeiro BRAZIL}

\author{and}

\author{Katia Cunha}
\affil{Observatorio Nacional, Rua General Jos\'e Cristino 77, CEP 20921-400,
Rio de Janeiro BRAZIL}
\bigskip

\author{Submitted to the Astrophysical Journal Letters}

\altaffiltext{1} {Based on observations made with the 2.1m telescope of
McDonald Observatory, University of Texas, USA}
\altaffiltext{2} {Based on observations made with the 1.52m telescope at the
European Southern Observatory under the agreement with the Observatorio
Nacional (Brazil)}

\begin{abstract}

We report on a high-resolution spectroscopic stellar parameter
and abundance analysis of 
a d$^{'}$ symbiotic star: the yellow component of StH$\alpha$190.  
This star has recently been discovered, and confirmed here, to be 
a rapidly rotating ($v\ sin i$ = 100$\pm$10 km s$^{-1}$)
subgiant, or giant, that exhibts radial-velocity variations of probably 
at least 40 km s$^{-1}$, indicating the presence of a companion (as in
many symbiotic systems, the companion is a hot white-dwarf star).
An analysis of the red spectrum reveals the cool stellar component to
have an effective temperature of T$_{\rm eff}$=5300$\pm$150K and a
surface gravity of log $g$ = 3.0$\pm$0.5 (this corresponds to an approximate
spectral type of G4III/IV).  These parameters result in an estimated primary
luminosity of 45L$_{\odot}$, implying a distance of about 780 pc (within
a factor of 2). 
The iron and calcium abundances are found to be close to solar, however,
barium is overabundant, relative to Fe and Ca, by about +0.5 dex.  
The Ba enhancement reflects mass-transfer of s-process enriched material
when the current white dwarf was an asymptotic giant branch (AGB) star,
of large physical dimension ($\ge$ 1AU).
The past and future evolution of this binary system depends critically
on its current orbital period, which is not yet known.
Concerted and frequent radial-velocity meassurements are needed to
provide crucial physical constraints to this d' symbiotic system.

\end{abstract}

\keywords{binaries: individual (StH$\alpha$190)---binaries: symbiotic---
stars: abundances---stars: rotation}

\section{INTRODUCTION}

The object StH$\alpha$190 was discovered during an objective-prism survey
for emission-line objects between -25$^\circ$ and 10$^\circ$ from the
galactic plane by Stephenson (1986), who classified it as a symbiotic
star.  It is known that symbiotic stars are a class of interacting 
binary systems consisting of a cool, evolved star and a hotter
companion star, with periods ranging from 1-3 years, up to 10-100
years (e.g. Allen 1982).  Based on infrared 
colors, symbiotics can be divided into three groups: s-, d-, and
d$^{'}$-types (Webster \& Allen 1975; Allen 1982).  Those that show 
dust continuum emission between wavelengths
of 1.0 and 5.0$\mu$m are classified as d-type and these systems
contain a mass-losing Mira variable as a cool component. Those with a
normal stellar spectrum in the infrared are classified as s-types.
Typically, the shorter period symbiotics are found among the s-types
while the longer period systems are found among the d-types. 
The possible binary properties of the d$^{'}$-types are less known; for
example, their binary periods (assuming they are binaries, as are 
the d- and s-types) are completely unknown.  The d$^{'}$ stars also 
exhibit infrared dust emission,
however, their dust color-temperatures are significantly lower
than in the d-types.  A more complete description about the known nature 
and properties of d$^{'}$-types is given in Allen (1982) and 
Schmid \& Nussbaumer (1993).

Schmid \& Nussbaumer (1993)
studied both optical and ultraviolet low-resolution
spectra of StH$\alpha$190 and classified it as a d$^{'}$-type symbiotic. 
More recently, Munari et al. (2001) have discovered that the cool component
of StH$\alpha$190  is a rapidly rotating G-giant (with 
$v\ sin i$$\sim$ 105 km s$^{-1}$).
They also found that the system exhibits high-speed, bipolar mass outflow. 
In this $Letter$ we report on an optical high-resolution
spectroscopic analysis of StH$\alpha$190, with the principal goal being 
to derive 
basic parameters (effective temperature, surface gravity, chemical 
composition, evolutionary state, and radial velocity) of this
symbiotic star.

\section{OBSERVATIONS}

The high-resolution spectra of StH$\alpha$190 analyzed here were obtained
on two different telescopes. Four spectra were taken
with the 2.1m telescope of McDonald Observatory with the Cassegrain 
cross-dispersed 
Sandiford echelle spectrometer plus CCD detector (McCarthy et al. 1993).
In addition, five high-resolution spectra were obtained on the
1.52m telescope at the European Southern Observatory (ESO) with the
fiber-fed coud\'e cross-dispersed echelle spectrometer FEROS and CCD 
detector (Kaufer et al. 1999).
The McDonald spectra were taken at a two-pixel
nominal resolving power of R$\approx$60,000 and covering wavelengths from
6100\AA\ to 7900\AA, while the ESO FEROS
spectrum has R=48,000, and with wavelength coverage from $\sim$4000-9000\AA.
A summary of all observations is shown in Table 1.

All McDonald spectra were reduced in a standard way with the IRAF software
package. The echelle frames were background corrected and flat-fielded.
The ESO spectra were reduced using the online pipeline reduction 
(Kaufer et al. 2000). 
Final S/N ratios were evaluated by the measurement of the rms flux
fluctuations, with values ranging from S/N= 80-150 in a continuum region 
at 6110\AA.  Spectra were checked for wavelength stability
by using telluric O$_{2}$ lines near 6275\AA\ as a reference and wavelength
differences were less than 6m\AA\ ($\sim$ 0.3 km s$^{-1}$) from spectrum
to spectrum.

\section{ANALYSIS AND DISCUSSION}

\subsection{The Rotation of the Yellow Primary of StH$\alpha$190}

Perhaps the most surprising aspect of StH$\alpha$190
is the presence of large numbers of absorption features broadened 
by a rather large rotational velocity (see also Munari et al. 2001).  
The wavelengths and strengths 
of many of these features coincide with absorption lines 
observed in late-F to early-K stars, e.g., Fe\,{\sc i} or Ca\,{\sc i}.
For stars within this spectral range, high-resolution spectra can often be
used to derive stellar parameters, 
as well as fairly detailed chemical compositions for
a range of chemical elements.  Due to the high rotational velocity of
StH$\alpha$190, however, the obtainable spectral resolution is   
degraded, thus preventing an analysis based on measurements 
of individual equivalent widths.  Instead, spectrum synthesis
techniques must be used, as it is difficult to isolate features
produced by only one species.  Because of the difficulty in analyzing
a range of different elements, in this {\it Letter} we concentrate on 
a small number of species (Fe\,{\sc i}, Fe\,{\sc ii}, Ca\,{\sc i}, and
Ba\,{\sc ii}) to derive stellar parameters (T$_{\rm eff}$ and log g)
and rough chemical abundances in order to place constraints on the 
StH$\alpha$190 system.
We confine a detailed analysis of selected absorption features
to the red part of the spectrum for two reasons.  First, the absorption
line-density decreases considerably from blue to red, allowing for a
clearer analysis of, as much as possible, nearly unblended spectral lines.  
Second, the possibility
of light contamination (or veiling) from a hot companion, or hot gas in
the system, diminishes towards the red.  In particular, 
Schmid \& Nussbaumer (1993)
discussed light contamination of the cool component's spectrum in
StH$\alpha$190 and concluded that there was some contamination
in the B-band, but very little by the V-band. 

The rotationally broadened lines of StH$\alpha$190 can be seen in the two 
panels of Figure 1, where sample 
spectral regions from the ESO spectrum taken on Oct. 27, 1999 are displayed. 
The regions shown
near 6150\AA\ and 6430\AA\ (as well as another region not illustrated near
5850\AA) were selected to be synthesized because they contain a number of   
broadened features whose respective strengths are dominated approximately
by single species, such as Fe\,{\sc i}, Fe\,{\sc ii}, Ca\,{\sc i}, and
Ba\,{\sc ii}. 
Synthetic spectra covering the regions shown in Figure 1 (the smooth
curves) were computed
using a recent version of the LTE synthesis code MOOG (Sneden 1973), a Kurucz
\& Bell (1995) linelist with solar gf-values, and Kurucz ATLAS9 model
atmospheres. 
The best-fit rotational velocity was determined by least-squares comparisons
between observed and synthetic spectra. The top panel of Figure 2 shows 
values of $\chi^{2}$ versus $v \sin i$. 
The lowest value found for $\chi^{2}$ is used to normalize the larger
values.  It is clear from Figure 2 that there is a sharp minimum near
$v \sin i$= 100$\pm$10 km s$^{-1}$.

This fact can already be used to provide some
constraints on stellar parameters if the assumption is made that the star
is rotating at less than its critical breakup velocity, where
\begin{equation}
V_{\rm critical} = \sqrt{\frac{GM}{R}} = [GMg]^{1/4}, 
\end{equation}
with $M$ being the stellar mass, $R$ its radius, and $g$ 
its surface gravitational
acceleration.  The bottom panel of Figure 2 shows this critical velocity
as a function of stellar mass for a series of surface gravities (shown as
log g), as well as the projected rotational velocity of StH$\alpha$190, 
shown as the
dashed line with arrows at either end.  The arrows indicate that this is a
lower value for the rotational velocity of StH$\alpha$190, which could 
be considerably
larger depending on the inclination angle.  It is clear from the bottom
panel of Figure 2 that, for any reasonable stellar mass, the 
surface gravity of StH$\alpha$190 must be larger than about log g $\sim$ 1.5,
or the star would be rotating in excess of its critical velocity. 

\subsection{Stellar Parameters, Chemical Composition, \& the 
Barium-Symbiotic Connection}

The atmospheric parameters, effective
temperature (T$_{\rm eff}$) and surface gravity (log $g$),
along with [Fe/H], [Ca/H], and [Ba/H] (where
[X/H]=log(N(X)/N(H))$_{\star}$- log(N(X)/N(H))$_{\odot}$)
abundances were determined using spectrum synthesis.
The solution for T$_{\rm eff}$ and log g
was obtained from a selection of 7 features near 6150\AA, or 6430\AA\ (and 
illustrated in Figure 1), as well as 5850\AA, where a single species 
dominated so that they
behaved almost as ``single'' spectral
lines.  Synthetic spectra covering a range in T$_{\rm eff}$ (from 4500 -
6500K), log g (from 2.0 to 4.0), and microturbulent velocity (from
$\xi$= 1.0 to 3.0 km s$^{-1}$) were computed and then compared to the
observed spectra.  A least-squares comparison was used to select the
best overall fits, with sample results illustrated for one particular set of
models shown in the upper panel of Figure 3: results from a set of models
with log g= 3.0 and $\xi$= 2.0 km s$^{-1}$ are presented for various effective
temperatures.  A set of models with fixed gravity and microturbulence
is shown because the largest changes in the chemical
abundances are found for changes in T$_{\rm eff}$.  Note that in
this case, both Fe\,{\sc i} and Fe\,{\sc ii} yield similar abundances for
an effective temperature of 5300K.  In all observed-synthetic comparisons
over these parameter ranges, the Fe, Ca, and Ba abundances were allowed to
vary in order to maximize the best fits.  

Reinspection of Figure 1 shows
that the Fe\,{\sc ii} features are quite sensitive to surface gravity. 
Estimates of the uncertainties in the derived stellar parameters would
indicate that log g can be constrained to about $\pm$0.5 dex, while
T$_{\rm eff}$ is uncertain by about $\pm$150K.  We note that an earlier
estimate of effective temperature for this star was given by Schmid \&
Nussbaumer (1992), based upon the dereddened (B-V), as T$_{\rm eff}$= 5150K.
The T$_{\rm eff}$ derived here would suggest a spectral type of about
G4 (Schmid \& Kaler 1982), compared to Munari et al.'s estimate of G7
and M\"urset \& Schmid's estimate of G5.  Based on the Schmid \& Kaler
calibration, the various estimated spectral types correspond to a range 
in T$_{\rm eff}$ from 5025--5300K. 

It was found that the best
overall fits can be obtained, based upon the Fe\,{\sc ii} features, with a 
gravity near log g= 3.0 and $\xi$= 2.0 km s$^{-1}$ (and these are the values
used to construct the top panel of Figure 3).  These fits also provide
near-solar Fe and Ca abundances (Ca\,{\sc i} is not plotted as it 
follows closely Fe\,{\sc i}). 
Barium is an element produced primarily via s-process neutron-capture 
nucleosynthesis (e.g., Wallerstein et al. 1997).  Note that in all
models shown in the top panel of Figure 3, the Ba\,{\sc ii} 6141\AA\
and 5853\AA\ features provide [Ba/H] abundances considerably larger 
than [Fe/H] (or [Ca/H]).  These results indicate that the G-type 
component in StH$\alpha$190 is enriched in Ba, relative to Fe, with
[Ba/Fe]$\sim$ +0.5.  This enhancement in StH$\alpha$190 is most easily 
understood as resulting from mass transfer in the system when the 
white-dwarf member was a thermally-pulsing asymptotic 
giant branch (AGB) star, and was s-process enriched during
its third dredge-up phase of stellar evolution.  s-Process
enhancements have been found in other symbiotic systems,
such as AG Dra (Smith et al. 1996), He2-467 (Pereira et al. 1998),
or S32 and UKS-Cel (M\"urset \& Schmid 1999). 

With derived stellar parameters of T$_{\rm eff}$= 5300K and log g= 3.0,
the current primary of StH$\alpha$190 can be located in a log g --
T$_{\rm eff}$ plane and compared to evolutionary tracks from stellar models.
This comparison is shown in the bottom panel of Figure 3, with evolutionary 
tracks taken from Schaerer et al. (1993) and consisting of models 
computed with solar abundances.  The parameters derived here suggest
that the yellow component of StH$\alpha$190 consists of a star with 
M$\sim$ 2.5M$_{\odot}$ nearing the base of the red giant branch. 
The values of mass, T$_{\rm eff}$, and g can be used to estimate a 
luminosity of L$\sim$ 45L$_{\odot}$ for the cool component of StH$\alpha$, 
or M$_{\rm V}$$\sim$ 0.65.
Adopting the V-magnitude from Munari et al. (2001) of V=10.5 and 
E(B--V)=0.10 (Schmid \& Nussbaumer 1993), a distance of
$\sim$780 pc is derived.  Uncertainties in the stellar mass, gravity, and
T$_{\rm eff}$ lead to an uncertainty in this distance of about a factor of
two (d$\sim$ 350--1500 pc).  This distance 
in fair agreement with Munari et al.'s estimate of 575 pc.  

\subsection{Radial Velocities and the Binary Nature of StH$\alpha$190}

Radial velocities are also obtained from the comparisons of observed to
synthetic spectra for StH$\alpha$190, which provided fairly accurate
velocity estimates in spite of the line broadening.  With nine spectra 
obtained over the time interval from 1995 to 2001, nine photospheric radial
velocities are presented here for the G-star component in StH$\alpha$190.  
The heliocentric velocities are listed in Table 1 and cover a small
range from 0 km s$^{-1}$ to +15 km s$^{-1}$.  Munari et al. (2001)
have also published radial velocities that overlap the values reported
here and range over some 40 km s$^{-1}$ in amplitude.  Both sets of radial
velocities are on the same scale, as demonstrated by comparisons of
interstellar Na D-line component velocities published by Munari et al.
(2001) and those observed here (differences are, this study -- Munari et al.
= -0.1$\pm$0.3 km s$^{-1}$).
Taken together, the radial velocities indicate that StH$\alpha$190, 
like other symbiotics, is a binary,
but almost certainly a closer binary than the d- and s-types.  Munari
et al. (2001) suggest a possible period for StH$\alpha$190 of 171 days,
although such a period does not fit very well the radial velocities  
observed here.  A possible shorter period of $\sim$37--39 days may provide
a better fit to both sets of radial velocities, however, more measured
velocities, spanning time scales from a few days to a few weeks, will be
needed to determine an accurate period. 

A definitive orbital period is crucial in establishing the past, and future,
history of StH$\alpha$190.  Under the assumption that the primary is in 
synchronous rotation, a short orbital period would be required.  In such a 
case, the projected rotational velocity of $v \sin i$=100 km s$^{-1}$ yields
a maximum binary period of 4.2 days (derived from the mass estimate and
gravity, which yields a radius of 8.3R$_{\odot}$).  Using the critical 
rotational velocity
as an upper limit (v$\le$ 230 km s$^{-1}$ from the bottom panel of
Figure 2), then a lower limit to the binary period of 1.8 days is derived.   
In such a system, StH$\alpha$190 would have undergone common envelope
evolution when the current white dwarf was an AGB star.  The current 
evolutionary stage of the G-star suggests that it will soon ascend the
red-giant branch, at which point another common envelope phase would occur,
resulting in further orbital shrinkage.  
On the other hand, if the period is much longer ($\sim$1--6 months) then the
system has probably not undergone common envelope evolution and may not in
the future.  In this case, the rapid rotation of the G-star is probably
due to spin-up from the accretion of material from a massive AGB
wind (Jeffries \& Stevens 1996). 

With the few and scattered radial-velocity
determinations to date, differences of at least 15 km s$^{-1}$, and
probably as large as 40 km s$^{-1}$, are indicated.
Observing StH$\alpha$190 intensively and obtaining high-quality
spectra will allow for the determination
of the orbital period, and will be crucial in establishing
the past and future evolution of StH$\alpha$190.

\section{CONCLUSIONS}

The first high-resolution spectral analysis for stellar parameters and
abundances in a d'-type symbiotic has
revealed it to contain a rapidly rotating G-subgiant, or early giant, with 
T$_{\rm eff}$=5300$\pm$150K and surface gravity of log g=3.0$\pm$0.5.
Abundances of [Fe/H] and [Ca/H] indicate a near-solar
metallicity, while barium is enhanced, with [Ba/Fe]$\sim$ +0.5.  Thus,
StH$\alpha$190 is s-process enriched and probably accreted matter from its
companion, now a white dwarf, when it was an s-process enriched thermally
pulsing AGB star.  Measured radial-velocity variations already betray the 
presence of the companion.  If the rapid rotation of the cool component 
reflects synchronous rotation, then the current orbital period must be
between about 1.8 to 4.2 days.  If the orbital period is much longer than
this, then the rapid rotation of the G-subgiant/giant is due presumably to
the addition of angular momentum from a massive AGB wind (Jeffries \&
Stevens 1996).  The current orbital period of StH$\alpha$190 is not yet
well-constrained, so additional radial velocity measurements of this
system, and other d'-symbiotics in general, are warranted.

\acknowledgements
We thank the staff of McDonald Observatory for technical support.  This
research is supported in part by the National Science Foundation 
(AST99-87374).

\clearpage

\figcaption[Figure1.ps]{Two spectral regions in StH$\alpha$190 are shown,
displaying the rotational profiles.
Top panel:  the observed StH$\alpha$190 spectrum is plotted
along with three synthetic spectra (smooth curves) computed for three
different surface gravities (at T$_{\rm eff}$= 5300K), showing the sensitivity
of the Fe\,{\sc ii} (and Ba\,{\sc ii}) to log g.  Bottom panel: 
the observed spectrum is shown as open circles.  Three different
synthetic spectra are shown to illustrate both the sensitivity of the
Fe\,{\sc ii} to gravity, and the good agreement obtained between the
6140\AA\ and 6430\AA\ regions (which also agrees with analysis of a region
near 5850\AA). 
\label{fig1}}

\figcaption[Figure2.ps]{Top panel: a least-squares comparison of
observed - synthetic spectra for various values of $v \sin i$ to 
illustrate the derivation of the rotational velocity in StH$\alpha$190: a clear
minimum is found at 100 km s$^{-1}$ (the lowest value found for $\chi$$^{2}$
is used to normalize the other values).  Bottom panel: 
the critical break-up velocity as a function of stellar
mass for various values of stellar surface gravities.  The dashed horizontal
line with arrows shows $v \sin i$ for StH$\alpha$190 (the arrows indicate that
this is a lower limit to the true equatorial rotational velocity). 
\label{fig2}}

\figcaption[Figure3.ps]{Derivations of abundances and stellar parameters
are illustrated here.  Top panel: abundances of [Fe\,{\sc i}/H],
[Fe\,{\sc ii}/H], and [Ba\,{\sc ii}/H] versus T$_{\rm eff}$ for 
model atmospheres with log g= 3.0 and microturbulence set to 2.0
km s$^{-1}$.  Note that Fe\,{\sc i} and Fe\,{\sc ii} give the same
abundances for T$_{\rm eff}$= 5300K, and that [Ba\,{\sc ii}/H] is always
larger than [Fe/H].  
Bottom panel: the G-type star in StH$\alpha$190 is placed in a
T$_{\rm eff}$ - log g plane, along with stellar model evolutionary tracks
from Schaerer (1993).  This comparison indicates that the yellow component
of StH$\alpha$190 is an intermediate-mass subgiant about to begin its ascent of the
red giant branch. 
\label{fig3}}

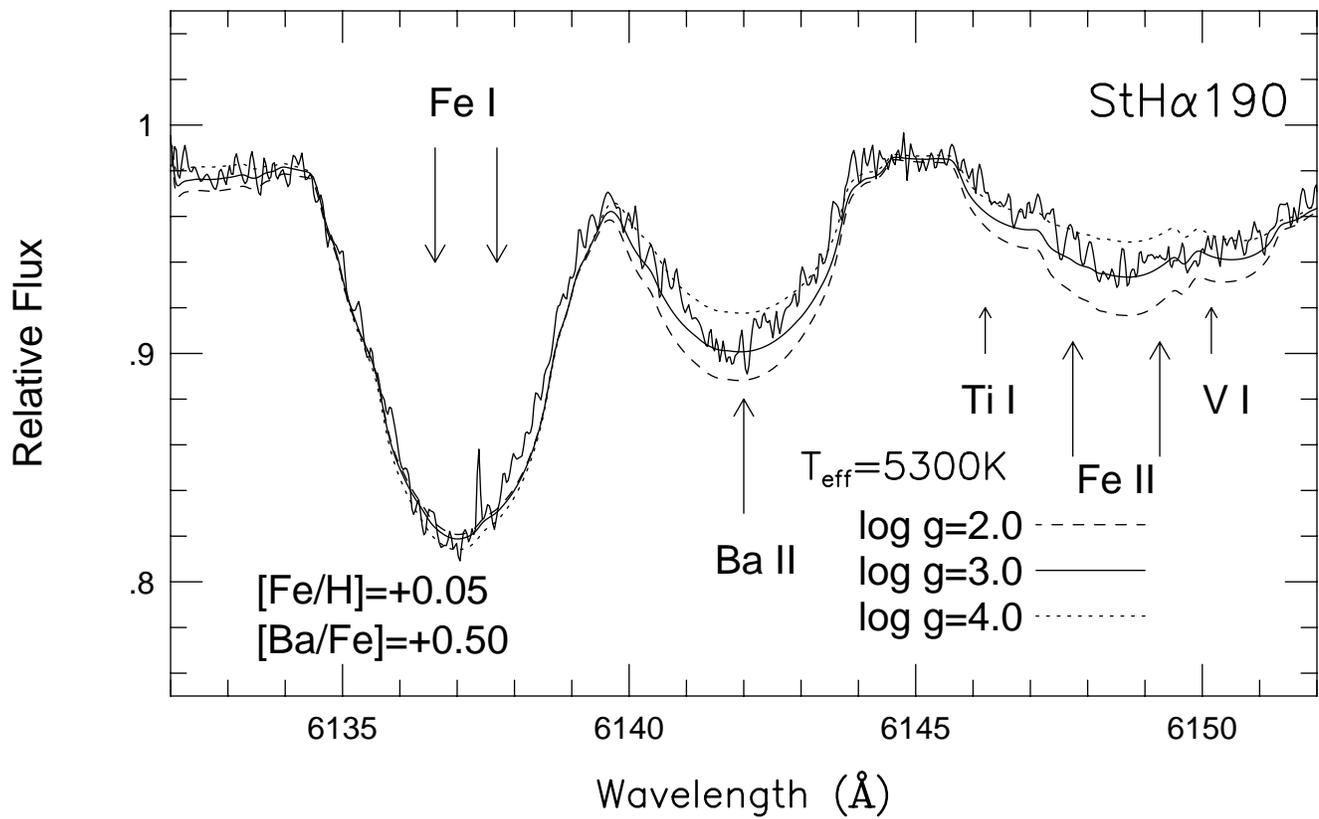
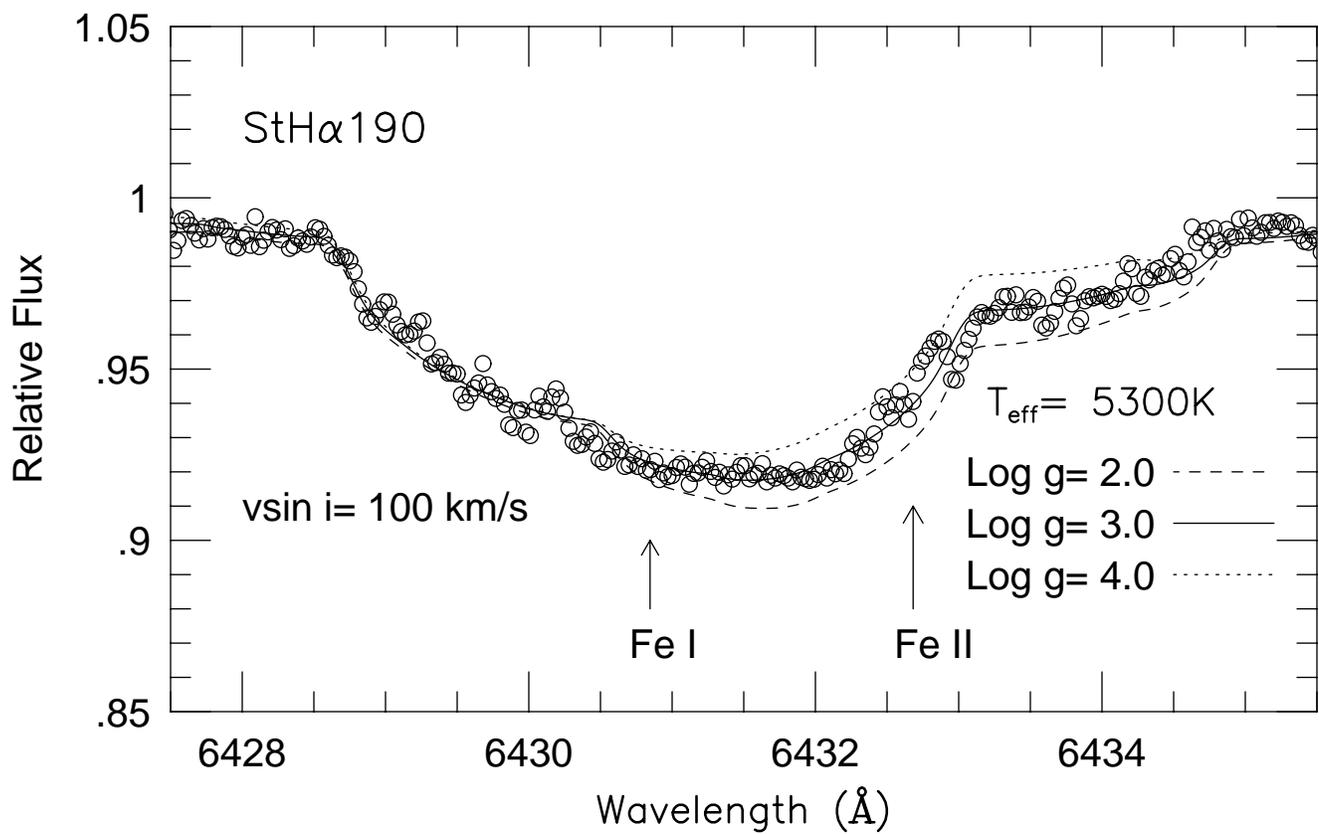
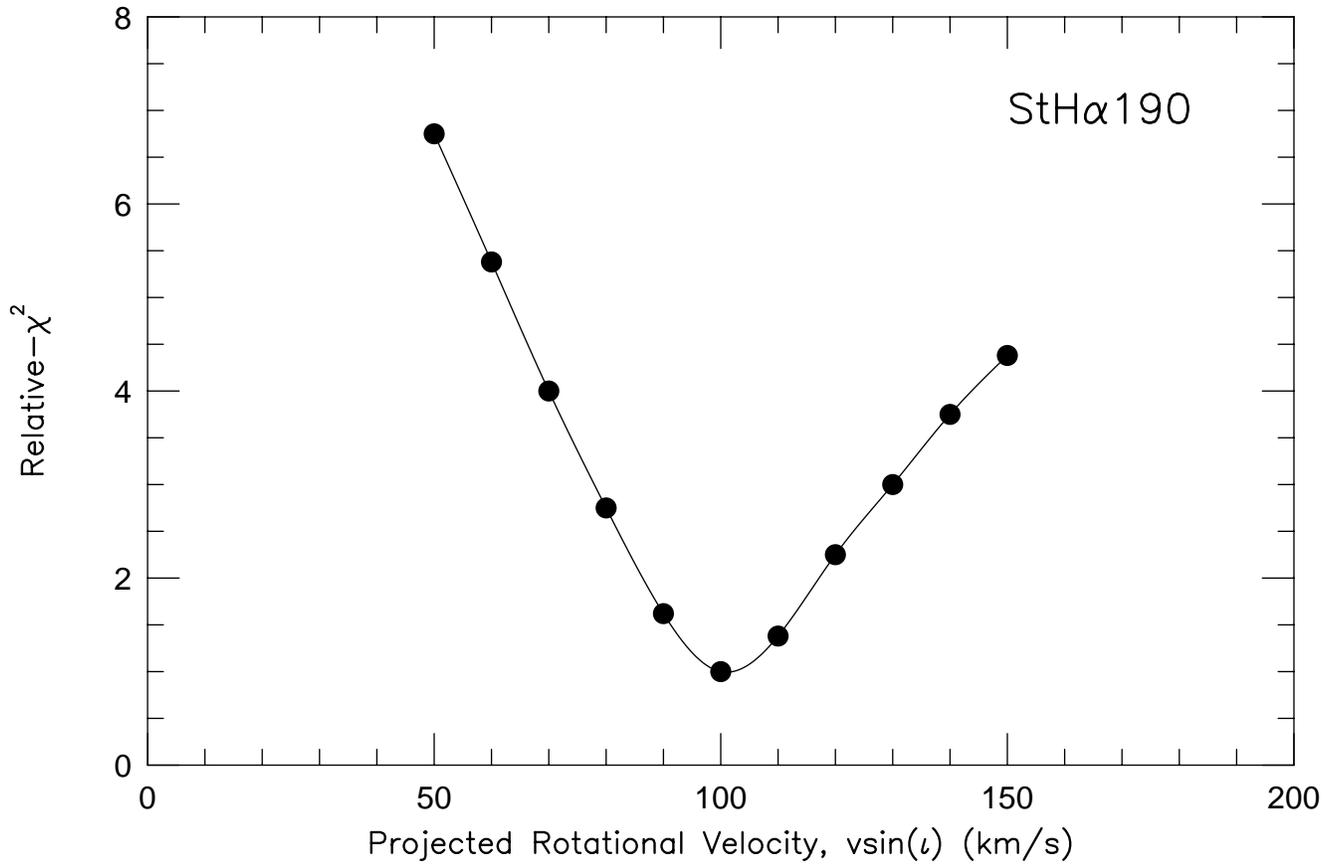
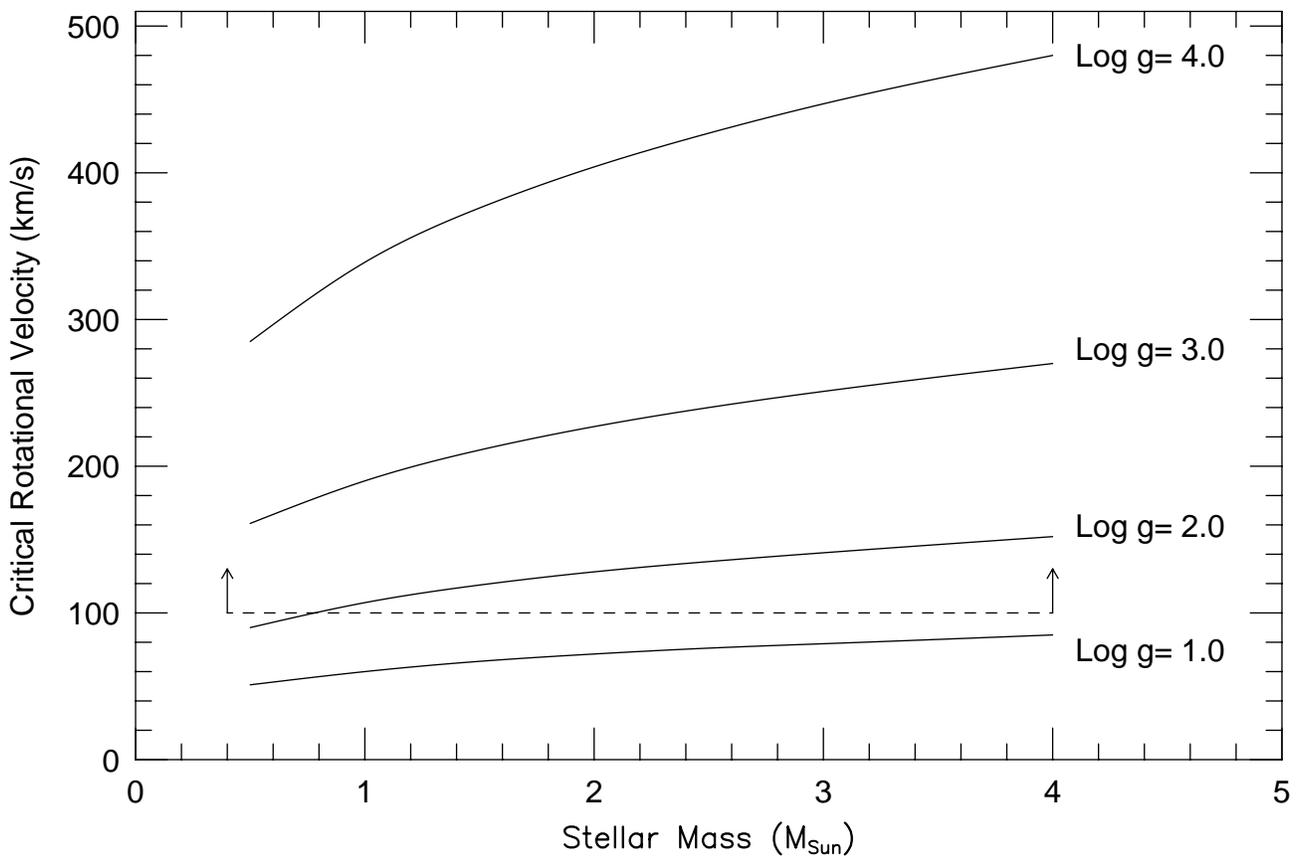
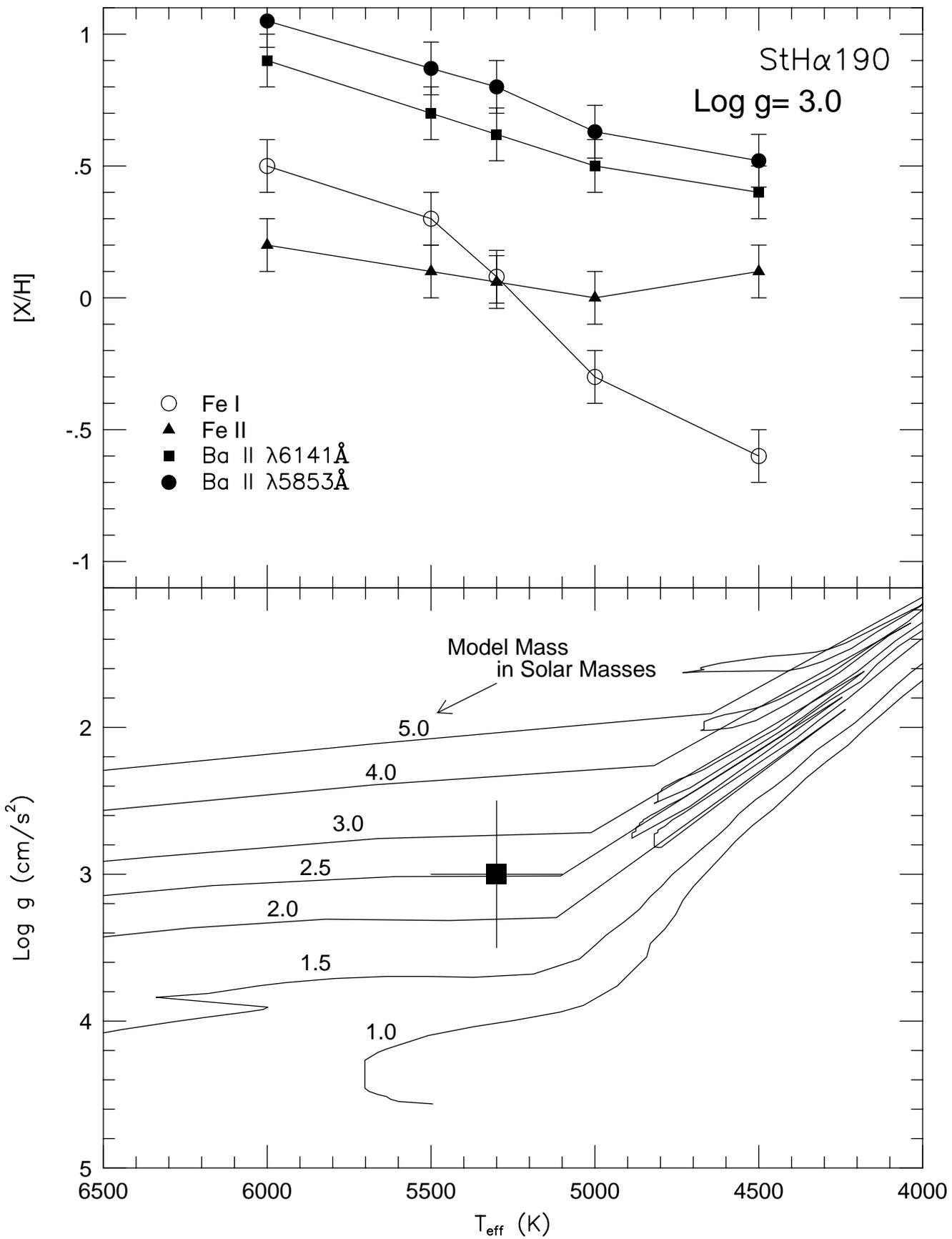
\end{document}